\documentclass[12pt]{article}
\begin{document}


\def\bkB{{\rm I\kern-.17em B}}
\def\B{\bkB}
\def\bkC{{\rm \kern.24em
         \vrule width.05em height1.4ex depth-.05ex
         \kern-.26em C}}
\def\C{\bkC}
\def\bkD{{\rm I\kern-.17em D}}
\def\D{\bkD}
\def\bkE{{\rm I\kern-.17em E}}
\def\E{\bkE}
\def\bkF{{\rm I\kern-.17em F}}
\def\F{\bkF}
\def\bkG{{\rm \kern.24em
         \vrule width.05em height1.4ex depth-.05ex
         \kern-.26em G}}
\def\G{\bkG}
\def\bkH{{\rm I\kern-.22em H}}
\def\H{\bkH}
\def\bkI{{\rm I\kern-.22em I}}
\def\I{\bkI}
\def\bkJ{{\rm \kern.19em
         \vrule width.02em height1.5ex depth0ex
         \kern-.20em J}}
\def\J{\bkJ}
\def\bkK{{\rm I\kern-.22em K}}
\def\K{\bkK}
\def\bkL{{\rm I\kern-.17em L}}
\def\L{\bkL}
\def\bkM{{\rm I\kern-.22em M}}
\def\M{\bkM}
\def\bkN{{\rm I\kern-.20em N}}
\def\N{\bkN}
\def\bkO{{\rm \kern.24em
         \vrule width.05em height1.4ex depth-.05ex
         \kern-.26em O}}
\def\O{\bkO}
\def\bkP{{\rm I\kern-.17em P}}
\def\P{\bkP}
\def\bkQ{{\rm \kern.24em
         \vrule width.05em height1.4ex depth-.05ex
         \kern-.26em Q}}
\def\Q{\bkQ}
\def\bkR{{\rm I\kern-.17em R}}
\def\R{\bkR}
\def\bkT{{\rm \kern.24em
         \vrule width.02em height1.5ex depth 0ex
         \kern-.27em T}}
\def\T{\bkT}
\def\bkU{{\rm \kern.30em
         \vrule width.02em height1.47ex depth-.05ex
         \kern-.32em U}}
\def\U{\bkU}
\def\bkZ{{\rm Z\kern-.32em Z}}
\def\Z{\bkZ}

\def\BBox{\vrule height 0.5em width 0.6em depth 0em}
\def\bbox{\hfill\framebox(6,6)}

\newtheorem{theorem}{Theorem}           
\newtheorem{prop}[theorem]{Proposition}
\newtheorem{lema}[theorem]{Lemma}
\newtheorem{corol}[theorem]{Corollary}
\newtheorem{defi}[theorem]{Definition}
\newtheorem{ex}[theorem]{Example}
\newtheorem{exs}[theorem]{Examples}
\newtheorem{obs}[theorem]{Remark}
\newtheorem{obss}[theorem]{Remarks}
\def\ds{\displaystyle}
\def\ra{\,\rightarrow\,}
\def\st{\stackrel}
\def\theequation{\thesection.\arabic{equation}}
\newenvironment{dem}{\noindent{\bf Proof:\\}}{ \hfill \BBox}
\def\p{\partial}
\def\ra{\;\rightarrow\;}

\newfont{\elevenmib}{cmmib10 scaled\magstep1}%
\renewcommand{\theequation}{\arabic{section}.\arabic{equation}}
\newcommand{\tabtopsp}[1]{\vbox{\vbox to#1{}\vbox to12pt{}}}
\font\larl=cmr10 at 24pt
\renewcommand{\thesection}{\Roman{section}}

\newcommand{\Title}[1]{{\baselineskip=26pt \begin{center}
            \Large   \bf #1 \\ \ \\ \end{center}}}
\hspace*{2.13cm}%
\hspace*{0.7cm}%
\newcommand{\Author}{\begin{center}\large \bf
           R. Caseiro${}^{a,b}$   \end{center}}
\newcommand{\Address}{\begin{center}
            $^a$ Universidade de Coimbra,\ \
     Departamento de Matem\'atica\\
     3000 Coimbra, Portugal \\
     $^b$ Universit\'e de Paris 6, Laboratoire ``GSIB"\\
     125 Rue du Chevaleret, 75013, Paris, France\\
      \end{center}}
\baselineskip=20pt
\bigskip
\bigskip
\bigskip

\Title{Master Integrals, Superintegrability and Quadratic
Algebras}

\Author

\Address \vspace{1cm}

\begin{abstract}
In this paper we use a generalization of Oevel's theorem about
master symmetries to relate them with superintegrability and
quadratic algebras.
\end{abstract}
\bigskip
\bigskip
\bigskip

\section{Introduction} \label{intro} \setcounter{equation}{0}

In this article a general framework is built in terms of master
symmetries and recursion operator to provide superintegrability
and quadratic algebras.

This is applied to the isotropic harmonic oscillator, the rational
Calogero-Moser system and the ``Goldfish" model.

\section{Background}

Let $M$ be a differentiable ($C^\infty$) manifold of finite
dimension and $\Lambda$ a bivector (a 2-times contravariant
skew-symmetric tensor field) on $M$. Associated with $\Lambda$
there is a natural morphism ${\ds \Lambda^\sharp}$ from the
cotangent bundle  $T^*M$ into the tangent bundle  $TM$ defined,
for all $\alpha,\beta\in T^*M$, by
\begin{equation}\label{sharp}
  <\Lambda^\sharp(\alpha),\beta>=\Lambda(\alpha,\beta),
\end{equation}
where $<.,.>$ denotes the usual coupling between $1$-forms and
vector fields.

We also define a bilinear map from
 $C^\infty(M)\times C^\infty(M)$ into $C^\infty(M)$ by

\begin{equation}\label{Poissonbracket}
  \{f,g\}=\Lambda(df,dg),\;\; f,g\in C^\infty(M).
\end{equation}

Due to the properties of $\Lambda$, this bracket satisfies

\begin{description}
  \item[PB1] ${\ds \{f,g\}=-\{g,f\}}$\mbox{ \bf\small\em{skew-symmetry}}
  \item[PB2]${\ds \{fh,g\}=f\{h,g\}+\{f,h\}g}$ \mbox{ \bf\small\em{Leibniz rule}}
\end{description}

We say that $(M,\Lambda)$ is a {\em Poisson manifold}, and
$\Lambda$ is a {\em Poisson tensor}, if, in addition, the bracket
(\ref{Poissonbracket}) satisfies the {\em Jacobi identity}

\begin{description}
  \item[PB3] ${\ds \{f,\{g,h\}\}+\{g,\{h,f\}\}+\{h,\{f,g\}\}=0}$.
\end{description}
which is equivalent to the vanishing of the Schouten-Nijenhuis
bracket $[\Lambda,\Lambda]$.

We call a vector field $X$  an {\em infinitesimal Poisson
automorphism} if it satisfies
$$
X\{f,g\}=\{X(f),g\}+\{f,X(g)\}, \;\;\; f,g\in C^\infty(M).
$$

Given a differentiable function $H$ on $M$ the {\em Hamiltonian
vector field associated with $H$} is the vector field defined by
\begin{equation}\label{Hamvector}
  X_H(x)=\Lambda^\sharp(x)(dH(x)),\;\; x\in M.
\end{equation}
If $\Lambda$ is a Poisson tensor then
\begin{equation}\label{prophamVF}
  [X_f,X_g]=[\Lambda^\sharp(df),\Lambda^\sharp(dg)]=\Lambda^\sharp(d\{f,g\})=X_{\{f,g\}}
\end{equation}
that is, $\Lambda^\sharp$ is a Lie algebra homomorphism between
the Lie algebra of differentiable functions ${\ds
(C^\infty(M),\{.,.\}})$ and the Lie algebra of vector fields
$(A(M), [.,.])$.

An integral of motion of $H$, or of ${\ds X_H}$, is a
differentiable function $F$ such that
\begin{equation}\label{intmotion}
  \{H,F\}=X_H(F)=0,
\end{equation}
so is constant along the orbits of the Hamiltonian system $X_H$
\begin{equation}\label{Hamsystem}
  \frac{d\phi}{dt}=X_H(\phi).
\end{equation}


By a \emph{Nijenhuis operator} $R$ in a manifold $M$ we mean a
$(1,1)-$tensor  satifying, for all vector fields $Z$ in $M$,
\begin{equation}\label{Nijtorsion}
 {\mathcal{L}}_{R(Z)}R=R{\mathcal{L}}_{Z}R.
\end{equation}

The Nijenhuis operators transform closed 1-forms into closed
1-forms in the following sense

\begin{prop}\label{ClosedForm}
Let $R$ be a Nijenhuis operator and $\alpha$ a closed $1-$form
such that ${\ds \alpha_1={}^tR\alpha}$ is also closed then, for
all $i\in\N$, ${\ds \alpha_i={}^tR^i\alpha_1}$ (where ${\ds
{}^tR^i}$ means ${\ds i^{th}}$ iterates of ${\ds {}^tR}$) are
closed.
\end{prop}

By a conformal vector field of a tensor $W$ we mean a vector field
$Z$ such that ${\ds {\mathcal{L}}_{Z}W=cW}$, for some constant
$c\in\bkR$.

 With Oevel's Theorem \cite{oevel}, a Nijenhuis
operators helps to define new symmetries if a conformal vector
field is known.

\begin{theorem}[Oevel]\label{Oeveltheorem}
Let $R$ be a recursion operator of $X_0\in A(M)$, this means ${\ds
{\mathcal{L}}_{X}R=0}$, and $Z_0\in A(M)$ a conformal vector field
of $X_0$ and $R$ such that
$$
{\mathcal{L}}_{Z_0}X_0=\lambda X_0,\;\;\; {\mathcal{L}}_{Z_0}R=\mu
R,\;\; \lambda,\mu\in\bkR.
$$

If $R$ is also a Nijenhuis operator then, defining the sequences
${\ds X_n=R^nX_0}$ and $Z_n=R^nZ_0$, $n\in\bkN$, we have, for all
$n,m\in\bkN_0$
$$
{\mathcal{L}}_{Z_n}R=\mu R^{n+1},
$$
$$
[Z_n,Z_m]=\mu(m-n)Z_{n+m}
$$
and
$$
[Z_n,X_m]=(\lambda+m\mu)X_{n+m}.
$$
\end{theorem}

The $Z_i$'s are called {\em master symmetries} or {\em symmetries
of second order} of the  vector fields $X_j$ because ${\ds
[[Z_i,X_j],X_j]=0}$ but ${\ds [Z_i,X_j]\neq 0}$ and they help us
define new symmetries of the system. We call {\em master
integrals} to functions $G$ which may not be integrals of motion
of the system but they induce  an integral $X(G)$.


Now let us consider two linearly independent Poisson structures
$\Lambda_0$ and $\Lambda_1$ in $M$. We say that they are
compatible if $\Lambda_0+\Lambda_1$  is again a Poisson tensor.
The compatibility condition is equivalent to the vanishing of the
Schouten-Nijenhuis bracket $[\Lambda_0,\Lambda_1]$.

A vector field $X\in A(M)$ is said to be {\em bihamiltonian} if it
is Hamiltonian with respect to two independent compatible Poisson
tensors, that is, if there exist two functions $H,F\in
C^\infty(M)$ such that
\begin{equation}\label{bihamVF}
  X=\Lambda_0^\sharp(dH)=\Lambda_1^\sharp(dF).
\end{equation}

There is an important example of bihamiltonian manifold: the
\emph{Poisson-Nijenhuis manifold} \cite{magri2}. This manifold is
special because one of the Poisson tensors is obtain from the
other by means of a  {\em Nijenhuis operator}. For instance, a
bihamiltonian manifold $(M,\Lambda_0,\Lambda_1)$ such that the
Poisson tensor $\Lambda_0$ is non-singular (symplectic manifold),
is a Poisson-Nijenhuis manifold with Nijenhuis operator ${\ds
R=\Lambda_1^\sharp\Lambda_0^{\sharp-1}}$. If ${\ds
X_1=\Lambda_0^\sharp(dH_1)=\Lambda_1^\sharp(dH_0)}$ is a
bihamiltonian system then we may define a sequence of symmetries
of $X$, ${\ds X_i=R^{i-1}X_1}$ and, if the first cohomology group
is trivial,  a sequence of integrals of motion in involution ${\ds
(H_i)_{i\in\N}}$ such that ${\ds X_i=\Lambda_0^\sharp(dH_i)}$, ie
${\ds dH_i={}^tR^i(dH_0)}$, $i\in\N$.


\section{The superintegrability and the cubic algebra}

Let  $X$ be a vector field  on a manifold $M$ of dimension $n$. It
is called  {\em maximally superintegrable} if it possesses
 $n-1$ functionally independent first integrals.

There are several examples of maximally superintegrable systems,
some of them are shown in the last section.

The superintegrability may be a consequence of the existence of a
sufficient number of master integrals, as the next proposition
shows.

\begin{prop}\label{MS}
Let $X$ be a vector field on a manifold $M$ and $G,F\in
C^\infty(M)$ master integrals of $X$. Then the function
\begin{equation}\label{MS1}
  L=X(G)F-X(F)G
\end{equation}
is an integral of the vector field $X$.
\end{prop}

\begin{dem}
Just notice that
\begin{eqnarray*}
X(L) &=& X(X(G))F+X(G)X(F)-X(X(F))G-X(F)X(G)\\
&=& X(X(G))F-X(X(F))G=0,
\end{eqnarray*}
because that $X(F)$ and $X(G)$ are integrals of the system.
\end{dem}

\begin{obs}
If $X$ is a vector field on a manifold of dimension $2n$ and if
$n$ functionally independent master integrals are known
$G_1,\ldots,G_n$ then we can define the integrals of motion
$F_i=X(G_i)$ and $L_{i,j}=X(G_i)G_j-G_iX(G_j)$ which may provide
the superintegrability of the system if $2n-1$ of them are
functionally independent.
\end{obs}


\begin{theorem}\label{LieAlg}
Let $X$ be a  Poisson infinitesimal automorphism on a Poisson
manifold $(M,\{.,.\})$. Suppose there exist master integrals of
$X$, ${\ds G_i}$, such that ${\ds \{G_i,X(G_i),\, i\in
J\subset\bkN\}}$ is a basis of a Lie subalgebra of
$(C^\infty(M),\{.,.\})$

Then, for all $i,j\in\bkN$ the functions $X(G_i)$ and ${\ds
L_{G_i,G_j}=X(G_i)G_j-X(G_j)G_i}$ generate a cubic algebra for the
Poison bracket.
\end{theorem}

\begin{dem}
Once $\{G_i,X(G_i), i\in J\}$ generates a Lie algebra, for each
$i,j,k\in J$ there exist constants $a_{i,j}^k$, $b_{i,j}^k$ such
that
\begin{equation}\label{LdepGs}
  \{G_i,G_j\}=\sum_{k\in J}(a_{i,j}^k G_k+ b_{i,j}^k X(G_k)).
\end{equation}

Applying $X$ twice to the last equation we obtain
$$
  X(\{X(G_i),(G_j)\}+\{G_i,X(G_j)\})=X(\sum_{k\in J}a_{i,j}^k
  X(G_k)),
$$
so
$$
\{X(G_i),X(G_j)\}=0.
$$

Writing ${\ds \{G_i,X(G_j)\}=\sum_{k\in J}(c_{i,j}^k
X(G_k)+d_{i,j}^k G_k) }$, with $c_{i,j}^k$, $d_{i,j}^k$ constants,
and noticing that
$$
X(\{G_i,X(G_j)\})=\{X(G_i),X(G_j)\}=0,
$$
we have ${\ds \sum_{k\in J}d_{i,j}^k X(G_k)=0}$, which yields
$d_{i,j}^k=0$.

Thus
\begin{eqnarray*}
\{L_{G_i,G_j},X(G_k)\}&=& X(G_i)\{G_j,X(G_k)\}-X(G_j)\{G_i,X(G_k)\}\\
&=& \sum_l [c_{j,k}^lX(G_i)X(G_l)-c_{i,k}^l X(G_j)X(G_l)]
\end{eqnarray*}
and
\begin{eqnarray*}
\lefteqn{\{L_{G_i,G_j},L_{G_k,G_h}\}=}\\
&=&
X(G_k)[L_{\{G_i,G_h\},G_j}-L_{\{G_j,G_h\},G_i}] +X(G_h)[L_{\{G_j,G_k\},G_i}-L_{\{G_i,G_k\},G_j}]\\
& & +L_{G_j,G_k}\{G_i,X(G_h)\}+L_{G_i,G_h}\{G_j,X(G_k)\}\\
&&-L_{G_j,G_h}\{G_i,X(G_k)\}-L_{G_i,G_k}\{G_j,X(G_h)\}\\
&=& \sum_l [X(G_k)(a_{i,h}^lL_{G_l,G_j}+b_{i,h}^l
L_{X(G_l),G_j}-a_{j,h}^lL_{G_l,G_i}-b_{j,h}^lL_{X(G_l),G_i})\\
&& +X(G_h)(a_{j,k}^lL_{G_l,G_i}+b_{j,k}^l L_{X(G_l),G_i}-a_{i,k}^lL_{G_l,G_j}-b_{i,k}^l L_{X(G_l),G_j})\\
&& +c_{i,h}^lL_{G_j,G_k}X(G_l)+c_{j,k}^lL_{G_i,G_h}X(G_l)
-c_{i,k}^lL_{G_j,G_h}X(G_l)-c_{j,h}^lL_{G_i,G_k}X(G_l)]\\
&=&\sum_l [X(G_k)(a_{i,h}^lL_{G_l,G_j}-a_{j,h}^lL_{G_l,G_i})+X(G_h)(a_{j,k}^lL_{G_l,G_i}-a_{i,k}^lL_{G_l,G_j})\\
&& +c_{i,h}^lL_{G_j,G_k}X(G_l)+c_{j,k}^lL_{G_i,G_h}X(G_l)
-c_{i,k}^lL_{G_j,G_h}X(G_l)-c_{j,h}^lL_{G_i,G_k}X(G_l)\\
 &&
+(b_{i,h}^lX(G_j)-b_{j,h}^lX(G_i))X(G_l)X(G_k)+(b_{j,k}^lX(G_i)-b_{i,k}^lX(G_j))X(G_l)X(G_h)]
\end{eqnarray*}

This last expression being cubic in the quantities $X(G_i)$ and
$L_{G_i,G_j}$, we refer to this result as that they generate a
cubic algebra.

\end{dem}

\begin{corol}
If  for each $i,j\in\N$, $\sum_l
b_{i,j}^lX(G_l)=b_iX(G_j)-b_jX(G_i)$ or all the cons\-tants
$b_{i,j}^k$ are zero then the integrals of motion genereate a
quadratic algebra.
\end{corol}

\begin{obs}\label{ConstGs}
Notice that in the above proposition we could have just demanded
that $d\{G_i,G_j\}$ be a linear combination of the $dG$'s and the
$dX(G)$'s.
\end{obs}


\begin{theorem}[Generalization of Oevel's theorem]\label{GenerOTHM}
Let $X$ be a vector field on a manifold $M$, $R$ a Nijenhuis
operator which is also a recursion operator of $X$ and $P$ a
$(1,1)-$tensor satisfying
\begin{equation}\label{ROcond1}
  {\mathcal{L}}_{X}P=a(R)
\end{equation}
and
\begin{equation}\label{ROcond2}
  {\mathcal{L}}_{PX}R=b(R),
\end{equation}
with $a(R)$, $b(R)$ polynomials in $R$. Then, defining the
sequences ${\ds X_i=R^iX}$, $Y_i=R^i(PX)$, $i\in\N_0$, we have
\begin{equation}\label{prop1}
[X_i,X_j]=0,
\end{equation}
\begin{equation}\label{prop2}
[X_i,Y_j]=a(R)(X_{i+j})-i b(R) (X_{i+j-1})
\end{equation}
\begin{equation}\label{prop3}
[Y_i,Y_j]=(j-i)b(R)Y_{i+j-1}.
\end{equation}
\end{theorem}

\begin{dem}
The proof is similar to the original Oevel's theorem proof, which
can be seen at \cite{oevel}.
\end{dem}

\

Suppose that $M$ has trivial first cohomology group and is endowed
with a non-de\-ge\-ne\-ra\-ted Poisson structure $\Lambda$ such
that ${\ds R\Lambda^\sharp=\Lambda^\sharp{}^tR}$ (this means that
the tensor $R\Lambda$ is a bivector). Futhermore suppose that
there exist functions such that ${\ds
X=\Lambda^\sharp(dH_1)=R\Lambda^\sharp(dH_0)}$ and ${\ds
Y=\Lambda^\sharp(dG_1)=R\Lambda^\sharp(dG_0)}$.


Then Proposition \ref{ClosedForm}  ensures us  that the $1-$forms
$$
\alpha_i={}^tR^i(dH_1),\;\;\; \beta_i={}^tR^i(dG_1),\; i\in\bkN
$$
are closed and we can consider them exact because of the
triviality of the first cohomology group.

Write ${\ds \alpha_i=dH_i}$ and ${\ds \beta_i=dG_i}$, for all
$i\in\bkN$.

First notice that $R\Lambda$ being a bivector yields
\begin{eqnarray*}
X_i(H_j)&=& <X_i,dH_j>= <X_{i+j}, dH_1> \\
&=& R^{i+j}\Lambda^\sharp( dH_1,dH_1)=0.
\end{eqnarray*}

Moreover (\ref{prop2}) ensures that the $G$'s are master integrals
of the $X$'s because
\begin{eqnarray*}
X_i(X_i(G_j))&=&X_i(\{H_i,G_j\})=-[X_i,Y_j](H_i)\\
&=& (i b(R) X_{i+j-1}-a(R)X_{i+j})(H_i)=0,
\end{eqnarray*}
relation (\ref{prop3}) implies
$$
d\{G_i,G_j\}=(j-i)b({}^tR)dG_{i+j-1}
$$
and
\begin{eqnarray*}
\{X_i(G_j),G_k\}&=& -Y_k(X_i(G_j)) =
[X_i,Y_k](G_j)-X_i(Y_k(G_j))\\
&=& i b(R) X_{i+k-1}(G_j)-a(R)X_{i+k}(G_j)-X_i(\{G_k,G_j\})\\
&=& i b(R)
X_{i}(G_{j+k-1})-a(R)X_{i}(G_{j+k})-(j-k)b(R)X_i(G_{k+j-1})\\
&=&(i+k-j) b(R) X_{i}(G_{j+k-1})-a(R)X_{i}(G_{j+k}).
\end{eqnarray*}
So ${\ds \{X_i(G_j),G_k\}}$ can be written as a linear combination
of the $X_i(G)$'s.

Now we can apply Theorem \ref{LieAlg}, with the $b$'s equal to
zero, and guarantee that, for each $i\in\N$ the integrals of
$X_i$, ${\ds X_i(G_j)}$ and ${\ds
L_{k,j}^i=X_i(G_k)G_j-X_i(G_j)G_k}$, $j,k\in\N_0$, close
quadratically under the Poisson bracket.

\

Futhermore notice that
$$
[R\Lambda,\Lambda](\Lambda^{\sharp-1}Y)={\mathcal{L}}_{RY}\Lambda+
({\mathcal{L}}_{Y}\Lambda)\circ{}^tR+({\mathcal{L}}_{Y}R)\circ\Lambda
$$
so, as $Y$ and $RY$ are Hamiltonian vector fields, we have
$$
[R\Lambda,\Lambda](\Lambda^{\sharp-1}Y)=({\mathcal{L}}_{Y}R)\circ\Lambda=b(R)\Lambda.
$$
Thus, if $R\Lambda$ is a Poisson tensor then it is compatible with
$\Lambda$, because $R$ is a Nijenhuis operator, and
$b(R)\Lambda=0$.

But this implies that
$$
b(R)X_i=b(R)\Lambda^\sharp(dH_i)=0 \mbox{ and }
b(R)Y_i=b(R)\Lambda^\sharp(dG_i)=0,
$$
so the relations (\ref{prop1}), (\ref{prop2}) and (\ref{prop3})
become
$$
[X_i,X_j]=[Y_i,Y_j]=0;\;\;\;\; [X_i,Y_j]=a(R)X_{i+j}
$$
and
$$
d\{G_i,G_j\}=0;\;\;\;\; d\{G_k,X_i(G_j)\}=a(R)X_i(G_{j+k}).
$$

\

Although in the last procedure  we need two master integrals,
Hamiltonians of the master symmetries,  to generate a new sequence
of integrals of motion, we may construct  it  only knowing one
master integral of all the vector fields.


\begin{prop}\label{masterIntAll}
Under the conditions of proposition \ref{GenerOTHM}, suppose there
exists a master integral $G$ of all the vector fields $X_i$, $i\in
N_0$, then the functions $G_i=Y_i(G)$, $i\in\N_0$ are also master
integrals of the same vector fields and, for each $k\in\N_0$,
$$
L_{i,j}^k=X_k(G_i)G_j-X_k(G_j)G_i, \mbox{ for all $i,j\in\bkN_0,
$}
$$
are integrals of $X_k$.
\end{prop}

\begin{dem}
Due to relation (\ref{prop2}), we have, for all $i,k\in \bkN_0$
\begin{eqnarray*}
X_k(G_i)&=& [X_k,Y_i](G)+Y_i(X_k(G))\\
&=& a(R)X_{i+k}(G)-k b(R)X_{i+k-1}(G)+Y_i(X_k(G)).
\end{eqnarray*}
But
$$
X_k(Y_i(X_k(G)))=a(R)X_{i+k}(X_k(G))-k
b(R)X_{i+k-1}(X_k(G))+Y_i(X_k(X_k(G)))=0,
$$
because $X_k(G)$ is an integral of all the $X_j$, so $Y_i(X_k(G))$
is an integral of $X_k$ and $X_k(G_i)$ is then an integral of
$X_k$. Thus $G_i$ is a master integral of all the $X_k$.

\end{dem}

\

\begin{obs}
It seems that in the previous procedure  only  one Hamiltonian
$G_0$ would be necessary, but note that the procedure applied to
$G_0$ yields $G_i=Y_i(G_0)=R^i\Lambda(dG_0,dG_0)=0$, i.e. all
master integrals are zero.
\end{obs}

\section{Examples }


\begin{ex}[Isotropic Harmonic Oscillator and the Fernandes' Theorem]

\

{\em The Iso\-tro\-pic Harmonic Oscillator is the Hamiltonian
system in ${\ds (\bkR^{2n},(q_i,p_i))}$, defined by the
Hamiltonian
\begin{equation}\label{HamOsc}
  {\mathcal{H}} = \sum_{i=1}^n
  \frac{1}{2}(p_i^2+q_i^2).
\end{equation}

It is well known that this system is completely integrable with
constants of motion in involution
\begin{equation}\label{HOintmotion}
  E_i=\frac{1}{2}(p_i^2+q_i^2),\;\;\; i=1,\ldots,n,
\end{equation}
and it is bi-Hamiltonian \cite{morosi} with respect to the
compatible Poisson tensors
\begin{equation}\label{HOpoisson0}
  \Lambda_0=\sum_{i=1}^n \frac{\p}{\p p_i}\wedge\frac{\p}{\p q_i}
\end{equation}
and
\begin{equation}\label{HOpoisson1}
  \Lambda_1=\sum_{i=1}^n E_i(\frac{\p}{\p p_i}\wedge \frac{\p}{\p
  q_i}).
\end{equation}

The Hamiltonian vector field can be expressed as
\begin{equation}\label{HOHamVF}
  X_{\mathcal{H}}=\sum_{i=1}^n(p_i\frac{\p}{\p q_i}-q_i\frac{\p}{\p
  p_i})=\Lambda_0^\sharp(d{\mathcal{H}})=\Lambda_1^\sharp(dH_0),
\end{equation}
with ${\ds H_0=\ln(E_1)+\ldots+\ln(E_n)}$. Defining the recursion
operator of the system
\begin{equation}\label{HOROp}
  R=\Lambda_1^\sharp\Lambda_0^{\sharp-1}=
  \sum_{i=1}^n E_i(\frac{\p}{\p q_i}\otimes dq_i+\frac{\p}{\p p_i}\otimes
  dp_i),
\end{equation}
a sequence of Hamiltonians and of Hamiltonian vector fields can be
defined
\begin{equation}\label{HOSeqHamVF}
  X_i=N^{i-1}X_{\mathcal{H}}=\Lambda_0^\sharp(dH_i), \;\;
  i=1,\ldots,n.
\end{equation}

Notice that the $(1,1)$-tensor
\begin{equation}\label{HOPop}
  P=\sum_{i=1}^n\varphi_i(\frac{\p}{\p q_i}\otimes dq_i+\frac{\p}{\p p_i}\otimes
  dp_i),
\end{equation}
with ${\ds \varphi_i=\arcsin(\frac{q_i}{\sqrt{q_i^2+p_i^2}})}$,
satisfies the conditions of Theorem \ref{GenerOTHM} with $a(R)=Id$
and $b(R)=0$. So defining the sequence

\begin{equation}\label{HOConfsym}
  Y_k=R^kPX=\sum_{i=1}^{n}\varphi_iE_i^k(p_i\frac{\p}{\p q_i}-q_i\frac{\p}{\p
  p_i}),
\end{equation}
we have
\begin{eqnarray*}
{[Y_i,Y_j]}& = & 0 \\
{[Y_i,X_{j}]} & = & -X_{i+j}
\end{eqnarray*}
and also
$$
 dZ_i(H_j)=0
$$
$$
[Y_i,R^j\Lambda_0]=-R^{i+j}\Lambda_{0}
$$
i.e., the vector fields $Y_i$ are not Hamiltonian with respect to
any of the Poisson structures.

But let us define the functions ${\ds G=\sum_{i=1}^n E_k
\varphi_k}$  and notice that
\begin{equation}\label{HOGfunct}
  X_i(G)=\sum_k E_k^{i-1}(p_k\frac{\p}{\p q_k}-q_k\frac{\p}{\p
  p_k})(G)=iH_{i}.
\end{equation}
So the functions ${\ds G_i=Y_i(G)}=\sum_k\varphi_k E_k^{i+1}$
satisfy
\begin{eqnarray*}
X_i(G_j)=X_{i+j}(G)=(i+j)H_{i+j},
\end{eqnarray*}
\begin{eqnarray*}
\{X_i(G_j),G_k\}&=&\{(i+j)H_{i+j},G_k\}\\
&=&(i+j)X_{i+j}(G_k)=(i+j)X_{i+j+k}(G)\\
&=&(i+j)(i+j+k)H_{i+j+k}
\end{eqnarray*}
and
\begin{eqnarray*}\label{HOGsfunctions}
 \lefteqn{\{G_i,G_j\}=\Lambda_0(dG_i,dG_j)=\Lambda_0(dZ_i(G),dZ_j(G))}\\
 &=&\Lambda_0(d(\sum_{k=1}^n
E_k^{i+1}\varphi_k), d(\sum_{h=1}^n
 E_h^{j+1}\varphi_h))=(i-j)Z_{i+j}(G)=(j-i)G_{i+j}.
\end{eqnarray*}

 Therefore, for each $k\in\N_0$ the integrals of motions of $X_k$, ${X_k(G_j)}$'s, and
$L^k_{i,j}=X_k(G_i)\,
G_j-X_k(G_j)\,G_i=(i+k)H_{i+k}G_j-(j+k)H_{j+k}G_i$ close
quadratically under the Poisson bracket defined by $\Lambda_0$.

\

Now let us consider a little more  general configuration, in which
the isotropic harmonic oscillator is a particular case.

\

Given a completely integrable Hamiltonian system $(M^{2n},\omega,
H)$ in a symplectic manifold,  Fernandes \cite{fernandes}
establishes necessary and sufficient conditions for the existence
of  a second Poisson structure giving the complete integrability
of the system, in a neighborhood of a fixed invariant torus.
Without lost of generality let us consider
\begin{equation}\label{manifold}
(M^{2n}=\bkR^n\times\bkT^n, (s_i,\theta_i)_{i=1}^n),\;\;
H=H(s_1,\ldots,s_n) \mbox{ and } \ds \omega=\sum_i ds_i\wedge
d\theta_i.
\end{equation}

\begin{defi}
Let $(x^1,\ldots,x^{n+1})$ be affine coordinates in a
$(n+1)$-dimensional affine space $A^{n+1}$. A hy\-per\-sur\-fa\-ce
in $A^{n+1}$ is called a {\em hy\-per\-sur\-fa\-ce of translation}
if it admits a pa\-ra\-me\-tri\-za\-ti\-on of the form
\begin{equation}\label{HT}
  (y^1,\ldots,y^n)\rightarrow
  x^j(y^1,\ldots,y^n)=a_1^j(y_1)+\ldots+a_n^j(y^n), \;\;\;
  (j=1,\ldots,n+1).
\end{equation}

\end{defi}

\begin{theorem}[\cite{fernandes}]
The completely integrable Hamiltonian system (\ref{manifold})
admits a second Poisson structure, giving its complete
integrability if and only if the graph of the Hamiltonian function
is a hypersurface of translation relative to the affine structure
determined by the action variables.
\end{theorem}

We present the ``only if" part of the proof because in what
follows the new Poisson structure will be needed.

\begin{dem}
Assume the $(M^{2n},\omega,H)$ is a completely integrable system
and that the graph of $H$ is a hypersuface of translation relative
to the action variables $(s^i)$, so it has a parametrization of
the form (\ref{HT}) with $x^i=s^i,\; i=1,\ldots,n$ and
$x^{n+1}=H$. We can choose the parameters $(y^i)$ so that the
Hamiltonian takes the simple form
$$
H(y^1,\ldots,y^n)=y^1+\ldots+y^n.
$$
If $(\varphi^1\ldots,\varphi^n)$ are the coordinates conjugated to
$(y^1,\ldots,y^n)$, we define a second Poisson structure by the
formula
$$
\Lambda_1=\sum_{i=1}^n y^i\frac{\p}{\p y^i}\wedge\frac{\p}{\p
\varphi^i}.
$$
One checks easily that the two Poisson structures are compatible,
and that the recursion operator is given by
$$
R=\sum_{i=1}^{n}y^i(\frac{\p}{\p y^i}\otimes dy^i + \frac{\p}{\p
\varphi^i} \otimes d\varphi^i).
$$
It is now clear from the expression of the Hamiltonian function in
the $y$-coordinates that ${\ds {\mathcal{L}}_{X_H}R=0}$, so the
vector field $X_H$ is bi-Hamiltonian.

\end{dem}

\

Thus a completely Hamiltonian system whose Hamiltonian's graph is
a hypersurface of translation relative to the affine structure
determined by the action variables, is bi-Hamiltonian
$$
X_H=\Lambda_0^\sharp(dH)=\Lambda_1^\sharp(dH_0)
$$
with ${\ds H_0=\sum_{i=1}^n\ln(y^i)}$.

The $(1,1)$-tensor
\begin{eqnarray*}
P=\sum_{i=1}^n(\varphi^i\frac{\p}{\p \varphi^i}\otimes d\varphi_i
+ \frac{\p}{\p dy_i}\otimes dy_i)
\end{eqnarray*}
satisfies
$$
{\mathcal{L}}_{X_H}P=Id \mbox{ and } {\mathcal{L}}_{PX}R=0
$$
so we can apply Theorem \ref{GenerOTHM} and conclude that the
 vector fields ${\ds Y_k=R^kPX=\sum_{i=1}^n y_k\varphi^i\frac{\p}{\p
\varphi^i}}$ and the function ${\ds G=\sum_{i=1}^n y^i \varphi^i}$
allows us to define the sequence of functions ${\ds G_i=Y_i(G)}$
($i=0,1,\ldots$), such that
$$
X_i(G)=iH_{i},
$$
$$
X_i(G_j)=X_{i+j}(G)=(i+j)H_{i+j},
$$
$$
\{X_i(G_j),G_k\}=(i+j)X_{i}(G_{k+j})
$$
and
$$
\{G_i,G_j\}_0=(i-j)G_{i+j}.
$$

Thus, for each $i\in\N$, the  integrals of motion of $X_i$,  ${\ds
L^i_{k,j}=X_i(G_k))G_j-X_i(G_j)G_k}$ and ${\ds X_i(G_j)}$, close
quadratically under the Poisson bracket defined by $\Lambda_0$. }
\end{ex}

\begin{ex}[The Rational Calogero-Moser System]

\

{\em

The rational Calogero-Moser system is an integrable Hamiltonian
system defined by

\begin{equation}\label{Hamiltoniano}
  {\cal H}=\sum_{i=1}^{n}(\frac{p_i^2}{2}+\frac{g^2}{2}\sum_{j\neq
  i}(q_i-q_j)^{-2}).
\end{equation}

\

It admits the pair of matrices $(L,M)$,
\begin{equation}\label{L}
  L_{ij}=p_i\delta_i^j + g\sqrt{-1}(q_i-q_j)^{-1}(1-\delta_i^j),
\end{equation}
\begin{equation}\label{M}
  M_{ij}=g\sqrt{-1}\sum_{h\neq i}(q_i-q_h)^{-2}\delta_i^j-g\sqrt{-1}(q_i-q_j)^{-2}(1-\delta_i^j)
\end{equation}
as a Lax pair.

 The Hamiltonian vector field with respect to the
canonical Poisson structure in $R^{2n}$ is

\begin{equation}\label{VF}
  X_H=\sum_i(p_i\frac{\partial}{\partial q_i}+\sum_{j\neq i}2(q_i-q_j)^{-3}\frac{\partial}{\partial
  p_i}).
\end{equation}

\

This system is completely integrable when we consider the sequence
of integrals of motion ${\ds F_i=Tr(L^i)}$, $i=1,\ldots n$.

Moreover, following \cite{Ranada}, if we consider the functions
${\ds G_i=Tr(QL^{i-1})}$, which provide the algebraic
linearization of the system \cite{cf},\cite{cfs}, the Hamiltonian
vector field becomes
\begin{equation}\label{HamVF}
  X_1=\sum_i F_i \frac{\p}{\p G_i}
\end{equation}
and  we may define the following compatible Poisson tensors
\cite{morosi}
\begin{equation}\label{CMpoisson0}
  \Lambda_0=\sum_i \frac{\p}{\p F_i}\wedge\frac{\p}{\p G_i},
\end{equation}
\begin{equation}\label{CMpoisson1}
  \Lambda_1=\sum_i F_i\frac{\p}{\p F_i}\wedge\frac{\p}{\p G_i}.
\end{equation}

The system $\dot{u}=X_1(u)$ is bi-Hamiltonian with respect to
these Poisson structures and the bi-Hamiltonian sequence of
integrals of motion is
\begin{equation}\label{CMchain}
  \Lambda_0^\sharp(dh_j)=\Lambda_1^\sharp(dh_{j-1}),\;\; j=0,1,\ldots
\end{equation}
where
$$
h_{-1}=\ln(F_1\ldots F_n),
$$
\begin{equation}\label{CMHgas}
  h_j=\frac{1}{2(j+1)}Tr(\Lambda_1^\sharp\Lambda_0^{\sharp-1})^{j+1}=\frac{1}{j+1}\sum_{i}(F_i)^{j+1},\;\;
  j=0,1,\ldots.
\end{equation}

Notice that ${\ds
X_1=\Lambda_0^\sharp(dh_1)=\Lambda_1^\sharp(dh_0)}$ and if we
define the sequence of Hamiltonian vector fields
\begin{equation}\label{CMseqVF}
  X_i=(\Lambda_1^\sharp\Lambda_0^{\sharp-1})^{i-1}X_1=\sum_k (F_k)^i\frac{\p}{\p G_k}\;\;
  i=1,2,\ldots,
\end{equation}
the following relation holds
\begin{equation}\label{CMrelation}
  X_i(\sum_k G_kF_k)=(i+1) h_i.
\end{equation}

Now consider the $(1,1)$-tensor
$$
P=\sum_{i=1}^n\frac{\p}{\p F_i}\otimes dG_i
$$
that satisfies
$${\mathcal{L}}_{PX}R=R \mbox{ and } {\mathcal{L}}_{X}P=Id$$
and define the sequence of master symmetries
\begin{equation}\label{CMseqconfsym}
  Y_i=(\Lambda_1^\sharp\Lambda_0^{\sharp-1})^iPX_1.
\end{equation}

Considering the functions $g_i=Z_i(\sum_k(G_k F_k))=\sum_k
F_k^{j+1}G_k$ we have
$$
X_i(g_j)=\sum_k F_k^{i+j+1},
$$
$$
\{X_i(g_j), g_k\}_0=(i+j+1)X_i(g_{j+k})
$$
and
$$
  \{g_i,g_j\}_0= (i-j)g_{i+j}.
$$

Now the Propositions \ref{LieAlg} and \ref{masterIntAll} ensure
that for each $X_i$, $i\in\N$, the integrals ${\ds X_i(g_j)}$ and
${\ds L^i_{k,j}=X_i(g_k)g_j-X_i(g_j)g_k}$ close quadratically
under $\{.,.\}_0$.

}
\end{ex}

\

\begin{ex}[The Goldfish System]

{\em

The Gold\-fish sys\-tem was first in\-tro\-du\-ced by
Ca\-lo\-ge\-ro \cite{calogero} and
 is the Hamiltonian system in $(\bkR^{2n},(q_i,p_i))$ defined by ${\ds X=\Lambda_0^\sharp(dH)}$,
 where $\Lambda_0$ is the canonical Poisson tensor and
\begin{equation}\label{HamGold}
  H=\sum_{i=1}^n\frac{g_i(q_i)}{\prod_{j\neq i}(q_i-q_j)}\mbox{\em{e}}^{a
  p_i},
\end{equation}
with $g_i$ arbitrary smooth functions, each one depending only on
the corresponding coordinate $q_i$ and $a$ an arbitrary constant.

\

Defining the Nijenhuis tensor
\begin{equation}\label{NT}
  R=q_i(\frac{\p}{\p q_i}\otimes dq_i+\frac{\p}{\p p_i}\otimes
  dp_i)
\end{equation}
and the function
\begin{equation}\label{Kapa}
  K=\sum_{i=1}^{^n}\frac{\rho_i g_i(q_i)}{\prod_{j\neq
  i}(q_i-q_j)}\mbox{\em e}^{a p_i},\;\; \rho_i=\prod_{j\neq i}q_j
\end{equation}
this system becomes  completely integrable and quasi-bihamiltonian
\cite{morosi}. The functions $c_i$, coefficients of the minimal
polynomial of $R$   ($q^n+\sum_ic_iq^{n-i}=\prod_{i=1}^n(q-q_i)$),
together with the integrals of motion
\begin{equation}
{\ds F_k=\sum_{i=1}^n\frac{\p c_k}{\p q_i}\frac{g_i(q_i) \mbox{\em
e}^{a p_i} }{\prod_{j\neq i}(q_i-q_j)}}
\end{equation}
 linearize algebraically the system because ${\ds \dot{c_k}=a
F_k}$.

In the coordinates $(c_i,F_i)$ the system has a simple
bi-Hamiltonian structure defined by the compatible Poisson tensors
\begin{equation}\label{bHStructure}
  Q_0=\sum_{i=1}^n\frac{\p}{\p c_i}\wedge \frac{\p}{\p F_i},\;\;
  Q_1=\sum_{i=1}^n F_i\frac{\p}{\p c_i}\wedge \frac{\p}{\p F_i}
\end{equation}
and the Hamiltonians
\begin{equation}\label{Hams}
  H_0=F_1+\ldots+F_n, \;\; H_1=\frac{1}{2}(F_1^2+\ldots+F_n^2).
\end{equation}

\

Similarly to last example, the $(1,1)$-tensor
$$
P=\sum_i\frac{\p}{\p F_i}\otimes dc_i
$$
and the function ${\ds G=\sum_{i=1}^n F_ic_i}$ allow us to define
the integrals of motion that close quadratically. }

\end{ex}

\section*{Acknowledgements}

The research reported in this article was made possible through
the agreement n$^o$ $502B2$ ``G\'eom\'etrie Diff\'erentielle et
Physique'' of the Luso-French cooperation ICCTI. Thanks are warmly
expressed to the Center of Mathematics of Coimbra's University, to
the GISB of the Universit\'e Pierre et Marie Curie, Paris VI, and
to the French Embassy in Lisboa.

\setcounter{equation}{0}


\end{document}